\documentclass[11pt]{article}
\usepackage{moriond,epsfig}

\def\be{\begin{equation}}
\def\ee{\end{equation}}
\def\bea{\begin{eqnarray}}
\def\eea{\end{eqnarray}}

\def\bar{\overline}

\def\b{\beta}

\def\e{\epsilon}

\def\bar{\overline}

\def\eV{{\rm eV}}
\def\ue3{\left| U_{e3} \right|}
\def\mnu{{\mathcal M}_{\nu f}}
\def\be{\begin{equation}}
\def\ee{\end{equation}}
\def\solm{\Delta_{\rm sun}}

\def\atm{\Delta_{\rm atm}}

\def\ord{{\mathcal O}}
\def\dt{\delta_\tau}
\begin{document}
\vspace*{4cm}
\title{Prediction of $U_{e3}$ and $\cos{2 \theta_{23}}$
from  discrete symmetry
\footnote{This talk is based on the work collaborated with
W. Grimus, A. S. Joshipura, S. Kaneko, L. Lavoura and H. Sawanaka \cite{NP}.}}

\author{Morimitsu Tanimoto}

\address{Department of Physics, Niigata University, 950-2181 Niigata, Japan}

\maketitle\abstracts{
We discuss the question why the mixing $U_{e3}$ is small.
The natural answer is  $U_{e3}=0$ in some symmetric limit, in which
two large mixings are realized.  
It is possible to force  $U_{e3}$ and $\cos{2 \theta_{23}}$ to be zero
by imposing a discrete symmetry.
We investigate a special class of symmetries $Z_2$
and of the consequences of their perturbative violation.}

\section{Introduction}

 In the standard model with three families, three mixing angles are
free parameters.  A lot of studies address the origin of the bi-large mixings
of neutrino flavors,
 which may be a clue to the beyond the standard model, on the other hand, 
we should also  answer the question 
why the  neutrino mixing $U_{e3}$ is so  small.

The natural answer is  $U_{e3}=0$ in some symmetric limit, in which
two large mixings are realized.  
There are many examples of symmetries
which can force $U_{e3}$ and/or $\cos{2 \theta_{23}}$ to vanish.
Both quantities vanish in the extensively studied
bi-maximal mixing \textit{Ansatz}~\cite{bd,nubim,lbim,sm},
which can be realized through a symmetry~\cite{nusmoh}. 
One can also make both $U_{e3}$ and $\cos{2 \theta_{23}}$ zero
while leaving the solar mixing angle arbitrary~\cite{mutau,d41}.
Alternatively,
it is possible to force only $U_{e3}$ to be zero
by imposing a discrete Abelian~\cite{low}
or  non-Abelian~\cite{d42} symmetry;
conversely,
one can obtain maximal atmospheric mixing but a free $U_{e3}$
 in a non-Abelian symmetry or a non-standard CP
symmetry~\cite{cp}. 

The symmetries mentioned above need not be exact.
It is important to consider perturbations of those symmetries
from the phenomenological point of view
and to study quantitatively~\cite{as}
the magnitudes of $U_{e3}$ and $\cos{2 \theta_{23}}$
possibly generated by such perturbations.
We discuss a special class of symmetries $Z_2$
and of the consequences of their perturbative violation.
We also study of the specific perturbation
which is induced by the electroweak radiative corrections
to a $Z_2$-invariant neutrino mass matrix
defined at a high scale.
The numerical result of a specific model is presented for this scenario.


\section{Vanishing $U_{e3}$ and  $Z_2$ symmetry}


Let us construct the neutrino mass matrix in terms of neutrino masses
 $m_1,m_2,m_3$ and mixings:
\begin{eqnarray}
 U = \left (\matrix{ c_{13} c_{12} & c_{13} s_{12} &  s_{13} e^{-i \delta}\cr 
  -c_{23}s_{12}-s_{23}s_{13}c_{12}e^{i \delta}
 & c_{23}c_{12}-s_{23}s_{13}s_{12}e^{i \delta} &   s_{23}c_{13} \cr
  s_{23}s_{12}-c_{23}s_{13}c_{12}e^{i \delta} 
& -s_{23}c_{12}-c_{23}s_{13}s_{12}e^{i \delta} & c_{23}c_{13} \cr}\right )\ ,
\label{parametrization}
\end{eqnarray}
\noindent where $c_{ij}$ and  $s_{ij}$ denote
 $\cos \theta_{ij}$ and $\sin \theta_{ij}$, respectively.
The neutrino mass matrix   $M_\nu$ is given in the flavor basis:
$M_\nu= U_{MNS}^* \ M_{\bf diagonal}  \ U_{MNS}^\dagger$.

In the standpoint of naturalness,
 as a dimensionless small parameter $s_{13}$ goes down to zero,
 the symmetry should be enhanced.
In $|U_{e3}|=s_{13}=0$ limit,  $M_\nu$ is written as  
\begin{eqnarray}
   M_\nu = \left (\matrix{\tilde X &  \tilde A &  \tilde B \cr 
                  \tilde A &  \tilde C &  \tilde D\cr 
                  \tilde B &  \tilde D &  \tilde E\cr}\right ) \ ,
\end{eqnarray}
where matrix elements are given including Majorana phases $\rho$ and $\sigma$:
\begin{eqnarray}
   &&\tilde X=c_{12}^2 m_1 e^{-2 i\rho}+s_{12}^2 m_2 e^{-2 i\sigma}  \ , \quad
\tilde A=c_{12}s_{12}c_{23}(m_2e^{-2 i\sigma}-m_1 e^{-2 i\rho})\ ,\nonumber\\
&&\tilde B=-c_{12} s_{12}s_{23}(m_2 e^{-2 i\sigma}-m_1 e^{-2 i\rho}) \ ,\quad
\tilde C=s_{12}^2c_{23}^2 m_1 e^{-2 i\rho}+
 c_{12}^2c_{23}^2m_2 e^{-2 i\sigma}+s_{23}^2 m_3 \ ,\\
 && \tilde D=c_{23}s_{23}(m_3-m_1  e^{-2 i\rho}s_{12}^2- 
m_2 e^{-2 i\sigma} c_{12}^2), \ 
\tilde E= s_{12}^2 s_{23}^2 m_1 e^{-2 i\rho}+
c_{12}^2 s_{23}^2 m_2 e^{-2 i\sigma}+c_{23}^2 m_3. \nonumber
\end{eqnarray}
\noindent
There is no explicit symmetry in this mass matrix 
 if there are no relations among matrix elements.
However, we obtain  the  mass matrix
with a $Z_2 $ symmetry by putting $\sin\theta_{23}=1/\sqrt{2}$:
\begin{eqnarray}
   M_{\nu f} = \left (\matrix{X &  A &  A\cr 
                  A &  B &  C\cr 
                  A &  C &  B\cr}\right ) \ ,
\label{matrix0}
\end{eqnarray}
with 
\begin{eqnarray}
 &&X=c_{12}^2 m_1  e^{-2 i\rho}+s_{12}^2 m_2  e^{-2 i\sigma} \ , 
 \ \  \qquad\qquad
A= -\frac{1}{\sqrt{2}}c_{12} s_{12}(m_1 e^{-2 i\rho} -m_2 e^{-2 i\sigma})\ ,
   \\
&&B=\frac{1}{2}(s_{12}^2m_1 e^{-2 i\rho}+c_{12}^2 m_2 e^{-2 i\sigma}+m_3)\ ,
\quad
C=\frac{1}{2}(s_{12}^2 m_1 e^{-2 i\rho}+c_{12}^2 m_2 e^{-2 i\sigma}-m_3) \ ,
\nonumber 
 \end{eqnarray}
where
\begin{eqnarray}
   S M_{\nu f} S =  M_\nu^{(0)} \ , \quad 
S =\left ( \matrix{1 &  0 &  0\cr 
                  0 &  0 &  1\cr 
                  0 &  1 &  0\cr}\right ), \quad S^2=1 \ .
\nonumber
\end{eqnarray}
The matrix $S$ is a realization of the discrete group $Z_2$.
It is emphasized that 
$m_1$, $m_2$, $m_3$, $\theta_{12}$, $\rho$, $\sigma$ are arbitrary.
In order to respect the symmetry, 
$\theta_{23}$ is  maximal, but  $\theta_{12}$ is not necessarily maximal.
  The general discussion of this symmetry 
was given in the previous work \cite{NP}.

\section{Non-zero $U_{e3}$,  $\cos 2\theta_{23}$ from $Z_{2}$ breaking}

Consider a general perturbation $\delta \mnu$ to  $M_{\nu f}$ 
in eq.(\ref{matrix0}).
The matrix $\delta \mnu$ is a general complex symmetric matrix,
but part of it can be absorbed through a redefinition
of the parameters in eq.(\ref{matrix0}).
The remaining part can be written,
without loss of generality,
as
\be
\delta \mnu = \left( \matrix{
0 & \e_1 & - \e_1 \cr
\e_1 & \e_2 & 0 \cr
- \e_1 & 0 & - \e_2 \cr}
\right).
\label{deltm}
\ee
The perturbation is controlled by two parameters,
$\e_1$ and $\e_2$,
which are complex and model-dependent.
We want to study their effects perturbatively,
i.e.\ we want to assume $\e_1$ and $\e_2$ to be small.
We define
two dimensionless parameters:
\be
\e_1 \equiv \epsilon A, \qquad \e_2 \equiv \epsilon^\prime B.
\ee
Thus, we have the neutrino mass matrix with $Z_2$ breaking as follows:
\be
\mnu = \left( \matrix{
X & A \left( 1 + \e \right) & A \left( 1 - \e \right) \cr
A \left( 1 + \e \right) & B \left( 1 + \e^\prime \right) & C \cr
A \left( 1 - \e \right) & C & B \left( 1 - \e^\prime \right) \cr}
\right) \ ,
\ee
\noindent where  we shall assume $\epsilon$ and $\epsilon^\prime$ to be small,
$\left| \e \right|, \left| \e^\prime \right| \ll 1$. 

One finds that,
to first order in $\e$ and $\e^\prime$,
the only effect of the $\delta \mnu$
is to generate non-zero $U_{e3}$ and $\cos{2 \theta_{23}}$.
The neutrino masses,
as well as the solar angle,
do not receive any corrections.
$U_{e3}$ and $\cos{2\theta_{23}}$ are of the same order as $\e$ and 
$\e^\prime$.
Define
\begin{eqnarray}
\hat m_1 \equiv m_1 e^{- 2 i \rho}\ , \quad 
\hat m_2 \equiv  m_2 e^{- 2 i \sigma}\ ,  \quad
\bar \e \equiv \left( \hat m_1 - \hat m_2 \right) \e\ , \quad 
\bar \e^\prime \equiv \frac
{\hat m_1 s_{12}^2 + \hat m_2 c_{12}^2 + m_3}{2} \ \e^\prime \ ,
\end{eqnarray}
we get
\begin{eqnarray}
\label{spue3}
U_{e3} &=&
\frac{s_{12} c_{12}}{m_3^2 - m_2^2} \left(
\bar \e s_{12}^2 \hat m_2^\ast
+ \bar \e^\ast s_{12}^2 m_3
- \bar \e^\prime \hat m_2^\ast
- {\bar \e^\prime}^\ast m_3
\right) \nonumber  \\  & &
+ \frac{s_{12} c_{12}}{m_3^2 - m_1^2} \left(
\bar \e c_{12}^2 \hat m_1^\ast + \bar \e^\ast c_{12}^2 m_3
+ \bar \e^\prime \hat m_1^\ast
+ {\bar \e^\prime}^\ast m_3
\right), 
\\
\cos{2 \theta_{23}} &=&
{\rm Re} \left\{
\frac{2 c_{12}^2}{m_3^2-m_2^2}
\left( \bar \e s_{12}^2 - \bar \e^\prime \right)
\left( \hat m_2 + m_3 \right)^\ast
- \frac{2 s_{12}^2}{m_3^2 - m_1^2}
\left( \bar \e c_{12}^2 + \bar \e^\prime \right)
\left( \hat m_1 + m_3 \right)^\ast
\right\}. \nonumber 
\end{eqnarray}

The induced values of $\ue3$ and $|\cos 2\theta_{23}|$ are strongly
correlated to neutrino mass hierarchies.
 This makes it possible to draw
some general conclusions even if we do not know the magnitudes of
$\e,\e'$. Remarks are given as follows:
\begin{itemize}
\item 
$U_{e3}$ gets
suppressed by a factor $\ord(\frac{\solm}{\atm})$ for the
inverted or quasi-degenerate spectrum with $\rho=\sigma=0$. Similar
suppression also occurs in the case of the normal neutrino mass hierarchy
even when $\rho\not =\sigma$. $U_{e3}$ need not  be suppressed in other
cases and can be large.

\item In contrast to $U_{e3}$, $\cos 2\theta_{23}$ is almost as large as 
$\e,\e'$ if neutrino mass spectrum is normal or inverted. It gets enhanced
compared to these parameters if the spectrum is quasi-degenerate.
\item
In case of the quasi-degenerate spectrum,
both $|\cos2\theta_{23}|$ and  $\ue3$  can become quite large and reach
the present experimental limits.
The parameters $\e,\e'$ are constrained to be
lower than $10^{-2}$ for the quasi-degenerate spectrum.
\end{itemize}

In our numerical study, the input parameters are   randomly varied
 in the experimentally allowed regions. $m_1$ was varied up to
$m_2$. On the other hand,  $\e,\e'$ are unknown 
unless the symmetry breaking is specified, so these are 
varied  in the range $-0.3\sim 0.3$ with the condition that the output
parameters should lie in the $90\%$ CL limit of the experimental data.

In Fig.1, we show the numerical result in the case of
 the normal neutrino mass hierarchy with $\rho=\sigma=0$.
The $\ue3$ is forced to be small less than $0.025$.
The value $\sim 0.025$ at the
upper end arises
from the (assumed) bound $|\e|,|\e'|\leq 0.3$. Since $\ue3$ is
proportional
to $\e,\e'$, it increases if  the bound on $\e,\e'$ is loosened.
However,  $|\e|\leq  0.3$ is a reasonable bound due to assume if
$Z_2$ breaking is perturbative. On the other hand,   
$|\cos 2\theta_{23}|$  can assume large values as seen from Fig. 1.
The present bound $\sin^2 2\theta_{23}>0.92$ from the atmospheric
experiments gets translated to $|\cos 2\theta_{23}|<0.28$ which constrains
 $|\e'| \leq 0.2$ in our analyses.
The phase dependence is found in the prediction of  $\ue3$, which 
 increases up to $0.075$ \cite{NP}. 


 We wish to point out an interesting aspect of
this analysis. Since $U_{e3}$ is zero in the absence of the perturbation,
the CP violating Dirac phase $\delta$ relevant for neutrino oscillations is
undefined at this stage. CP violation is present through the Majorana
phases $\rho$ and $\sigma$. Turning on perturbation leads to non-zero
$U_{e3}$ and also to a non-zero Dirac phase even if perturbation is real.
Moreover, $\delta$ generated this way can be large and independent of the
strength of perturbation parameters \cite{asn}.
\begin{figure}
\hskip 2 cm \psfig{figure=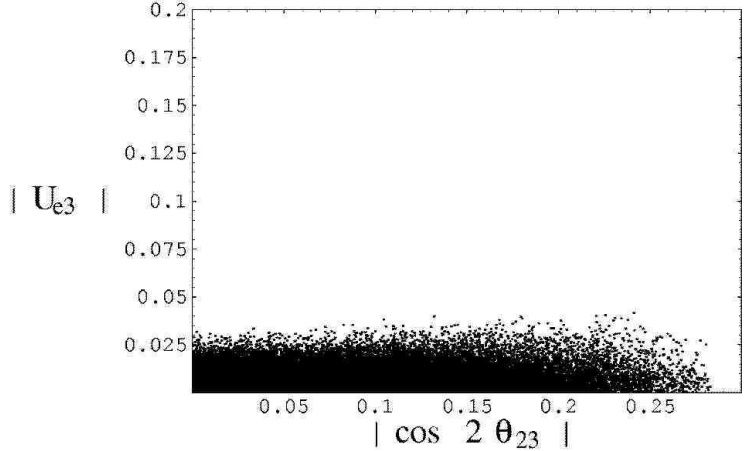,height=6 cm}
\caption{In case of the normal neutrino mass hierarchy with 
$\rho=0, \ \sigma=0$ and $\e,\e'=-0.3\sim 0.3$.}
\end{figure}


\section{Radiatively generated $U_{e3}$ and $\cos 2\theta_{23}$}

The $\e,\e'$ were treated as independent parameters so far. They can be
related in specific models. We now consider one example which is based
on the electroweak breaking of the $Z_2$ symmetry in the MSSM.
We assume that neutrino masses are generated at some
high scale $M_X$ and the effective neutrino mass operator describing them
is  $Z_2$ symmetric with the result that $U_{e3}= \cos 2\theta_{23}=0$ 
at $M_X$. This symmetry is assumed to be broken spontaneously
in the Yukawa couplings of the charged leptons. This breaking would
radiatively induce non-zero $U_{e3}$ and $\cos 2\theta_{23}$ \cite{Ratz}.
 This can be
calculated by using the renormalization group equations (RGEs) of the
effective neutrino mass operator \cite{rg1,rg2,rg3}. 
These equations depend upon the 
detailed structure of the model below $M_X$. We assume here that theory
below $M_X$ is the MSSM and use the RGEs
derived in this case. Subsequently we will give an example 
which realizes our assumptions.

Integration of the RGEs allows us \cite{rg1,rg2,rg3} to relate the
neutrino mass matrix $\mnu (M_X)$ to the corresponding matrix at the low scale
which we identify here with the $Z$ mass $M_Z$:
\be \label{corrected}\mnu(M_Z) \approx I_g
I_t ~(I~ \mnu (M_X) ~I~)\ , \ee
where $I_{g,t}$ are calculable numbers depending on the gauge and
top quark Yukawa couplings. $I$ is a flavor dependent matrix
given by
\be
 I\approx \mathrm{diag}(1+\delta_e,1+\delta_\mu,1+\delta_\tau)  \quad
 {\rm with} \quad
\delta_\alpha\approx c\left({m_\alpha\over 4 \pi v }\right)^2 \ln{M_X\over
M_Z}~, 
\ee
where $c=\frac{3}{2},-\frac{1}{\cos^2\b}$ in case of the SM
 and  the MSSM,
respectively \cite{rg1}. $v$ refers to the VEV for the SM Higgs doublet.

Since $\mnu(M_X)$ is given by eq.~(\ref{matrix0}), we can write 
$\mnu (M_Z)$ as follows when
the muon and the electron Yukawa couplings are neglected:
\begin{eqnarray}
\mnu (M_Z)
&=& \left(\matrix{ X&A'&A'\cr
                       A'&B'&C'\cr
                       A'&C'&B'\cr}\right)
    +\left( \matrix{ 0&A' \e&-A'\e \cr
                       A' \e&B' \e'&0 \cr
                       -A'\e&0&-B'\e' \cr} \right)+O(\delta_\tau^2) ~,
\end{eqnarray}          
where
\be 
C'=C(1+\dt)~~,~~A'=A(1+\frac{\dt}{2})~~,~~
B'=B(1+\dt)~~,~~\e=\frac{\e'}{2}=-\frac{\dt}{2} \ .
\ee
Note that $m_1$, $m_2$ and $m_3$ defined previously are no longer mass
eigenvalues
because of the changes $A\rightarrow A'$,  $B\rightarrow B'$ and
  $C\rightarrow C'$.
Then we get 
\begin{eqnarray}
U_{e3} \simeq&& \hskip -0.5cm -{ \dt s_{12}c_{12}\over2(
m_3^2-m_1^2)}\left[m_1^2+2 m_3 \hat{m_1}^*+m_3^2\right] 
 + { \dt s_{12}c_{12}\over 2 m_3^2-m_2^2}
\left[m_2^2+2 \hat{m_2}^* m_3+m_3^2 \right]\ , \nonumber \\ 
\cos 2 \theta_{23} \simeq&& \hskip -0.5cm  {\dt s_{12}^2\over
m_3^2-m_1^2}\left[m_1^2+2 m_3 \hat{m_1}^*+m_3^2\right]
+ {\dt c_{12}^2\over m_3^2-m_2^2}
\left[m_2^2+2 \hat{m_2}^* m_3+m_3^2 \right] \ .
\end{eqnarray}

It is seen that the effect of the radiative corrections is enhanced 
in the case of the  quasi-degenerate neutrino masses with  
$|\rho-\sigma|=\pi/2$ as  previous works presented \cite{rg2,rg3}.
In the MSSM, the parameter $\dt$ is negative and its absolute value 
 can become quite large for large $\tan\b$.
\begin{figure}
\hskip 2 cm \psfig{figure=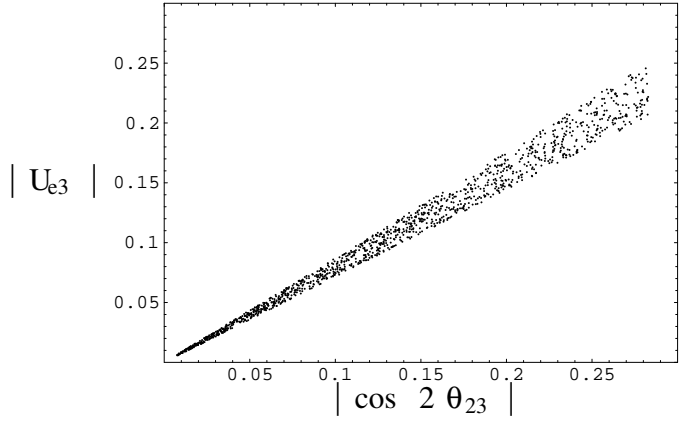,height=5.5 cm}
\caption{In case of the radiatively broken $Z_2$ and the quasi-degenerate 
neutrino masses with $\rho=0,\ \sigma=\pi/2$.}
\end{figure}
Results of the numerical analysis are shown in Fig. 2 in case of the
quasi-degenerate spectrum with $m=0.3 ~\eV, \sigma=\pi/2,\rho=0$. 
Both $\ue3$ and $|\cos 2\theta_{23}|$ can reach their respective 
experimental bound.  We find numerically
that $\tan\beta$ is constrained to be lower than $20$ in this case.
 On the other hand,  $\ue3$ reaches at most $0.025$
in the normal-hierarchy  and inverted-one.
  The forthcoming
experiments will be able to test this relationship between $\ue3$ and 
$|\cos 2\theta_{23}|$.
\section{Summary}
The neutrino mixing matrix contains two small parameters $\ue3$ and $\cos
2\theta_{23}$ which would influence the outcome of the future neutrino
experiments. 
The vanishing of $\ue3$ was shown to
follow from a class of $Z_2$ symmetries of ${\mathcal M}_{\nu f}$. 
This symmetry
can be used to parameterize all models with zero $U_{e3}$. A specific
$Z_2$ in this class also leads to the maximal atmospheric neutrino mixing
angle. We showed that breaking of this can be characterized by two
dimensionless parameters $\e,\e'$ and we studied their effects
perturbatively and numerically.

It was found that the magnitudes of $\ue3$ and  $|\cos 2\theta_{23}|$
are  strongly dependent upon the neutrino mass hierarchies
and CP violating phases. The $\ue3$ gets strongly suppressed in case of 
the inverted or quasi-degenerate neutrino spectrum if $\rho=\sigma$ while
similar suppression occurs in the case of normal hierarchy independent of 
this phase choice. The choice $\rho\not =\sigma$ can lead to a larger
values $\sim 0.1$ for $\ue3$ which could be close to the experimental
value in some cases with inverted or quasi-degenerate spectrum.
 In contrast, the $|\cos 2\theta_{23}|$ could be
large, near its present experimental limit in most cases studied. For the
normal and inverted mass spectrum, the magnitude of  $\cos 2\theta_{23}$
is similar to the magnitudes of the perturbations $\e,\e'$ while it can
get enhanced compared to them if the neutrino spectrum is quasi-degenerate.

\end{document}